\documentclass[onecolumn]{article}

\usepackage{graphicx}
\usepackage{subfigure}
\usepackage{amsmath}
\usepackage{amsfonts}
\usepackage{amssymb}
\usepackage{amsbsy}
\usepackage{times}
\usepackage[colorlinks=true,citecolor=black,linkcolor=black]{hyperref}
\usepackage{graphics}
\usepackage{graphicx}
\usepackage[]{subfigure}

 \usepackage{cite}
 \usepackage{anysize}\marginsize{10mm}{10mm}{10mm}{10mm}

\title{Lateral Stability Analysis of Hypersonic Vehicle under Pressure Fluctuation by Solving Mathieu Differential Equation}
\author{
Qingkai Wei,%
    \thanks{
Ph.D. student, Department of Aeronautics and Astronautics; weiqingkai@pku.edu.cn.}
  Xun Huang
  \thanks{
  Associate Professor. State Key Laboratory of Turbulence and Complex Systems, Department of Aeronautics and Astronautics, College of Engineering. Corresponding author: huangxun@pku.edu.cn.} \\
  {\normalsize\itshape
   Peking University, Beijing, 100871, People's Republic of China}\\
}

\begin{document}
\maketitle
\begin{abstract}
Two recent test failures of Hypersonic Technology Vehicle 2 impose a strike to the increasingly growing enthusiasm, not only on the United States side. It is important to find out the exact failure reason, otherwise a solution is impossible. In this Note, we propose a potential failure reason from the perspective of lateral stability analysis. We argue that the time variant pressure fluctuations, which are normally omitted in classical aircraft dynamics analysis, could not be neglected in dynamic analysis of hypersonic vehicles. To demonstrate the idea, a hypersonic model is imagined in this work and its aerodynamic parameters are estimated using fundamental fluid principles. Pressure fluctuations are thereafter estimated by an empirical formula. A lateral dynamic equation is set up, taking those time variant fluctuations into account. The resulted equation is a Mathieu differential equation. Numerical solutions of this equation show that the inclusion of fluctuation terms generates more complicated dynamics and should be considered in flight controller design.
\end{abstract}
%
%
\section{Introduction}
Hypersonic vehicle has become a popular topic once again in aerospace community\cite{Calise97,Wang00,Sachs05,Sigthorsson08,Mani09}. However, two recent test failures of Hypersonic Technology Vehicle (HTV-2) \cite{DARPA} impose a strike to the increasingly growing enthusiasm, not only on the United States side. It is important to find out the exact failure reason, otherwise a solution is impossible. In this work, we propose a potential failure reason from the perspective of lateral stability. We argue that the time variant pressure fluctuations, which are normally omitted in classical aircraft dynamics analysis, could generate trouble for hypersonic vehicle dynamics.  

Flight dynamics and control has been recognized as an important research topic since Wright brothers first flied to sky in 1903. Lateral and longitudinal stability are two classical flight dynamics and control problem. In the lateral stability problem, coupling dynamics was an interesting phenomenon that can produce either negative or beneficial stability for various aircraft design. 
According to the released information, \cite{DARPA} serious dynamic coupling between yaw and roll causes, at least, the first test failure of HTV-2. Similar coupling problems have been discovered in X-series aircrafts and space shuttle. A good summary of those NASA practices can be found in the literature\cite{day1997coupling}. 

For the HTV-2 case, some one may argue that a coupling between pitch and lateral dynamics could exist. In contrast, we hypothesize that the failure could still happen in a classical way, that is, the dynamic divergence develops in lateral coordinates. However, we should take new physical factors into account. It is well known that hypersonic flight generates huge acoustic pressure and suffers vibroacoustic problems. It is interesting to know if the acoustic wave, or more generally, the fluctuating pressure, could impede the dynamic stability. A methodology is introduced in this paper to analyze the effect of surface pressure fluctuations on lateral dynamic stability, by solving a so-called Mathieu differential equation. It is worthwhile to point out that the proposed method should be generic, although the initial idea comes from analyzing the HTV-2 case.

\section{Vehicle and Aerodynamic Information }
Pressure fluctuations in turbulent boundary layer \cite{blake1970turbulent} were mainly studied for vibroacoustics. Experimental investigations have been conducted at subsonic speed to measure surface pressure fluctuations on a smooth wall, which could induce panel fluttering and structural vibrations.

\begin{figure}
     \centering
     \subfigure[]
       {\includegraphics[height=35mm]{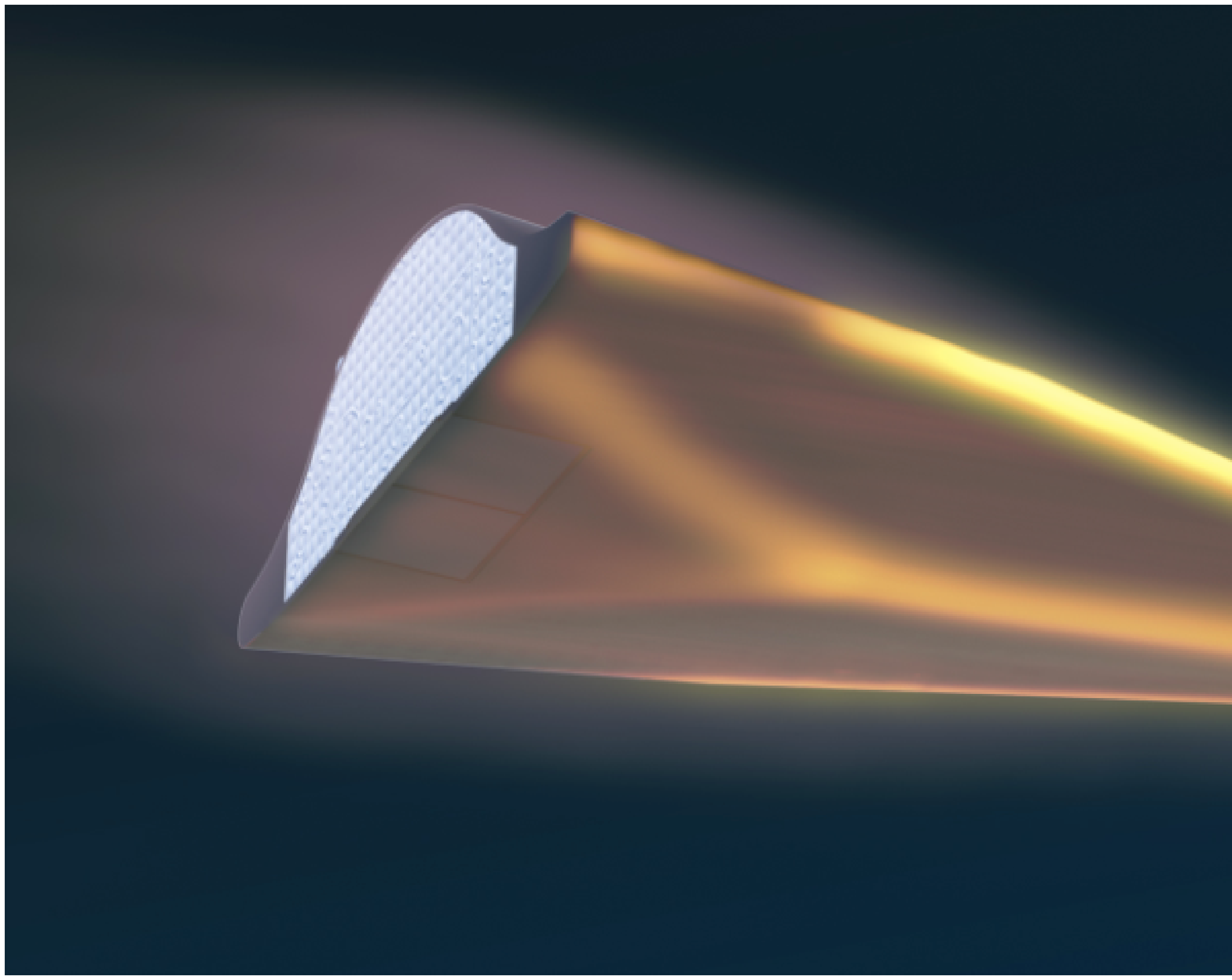}}
     \subfigure[]
       {\includegraphics[height=35mm]{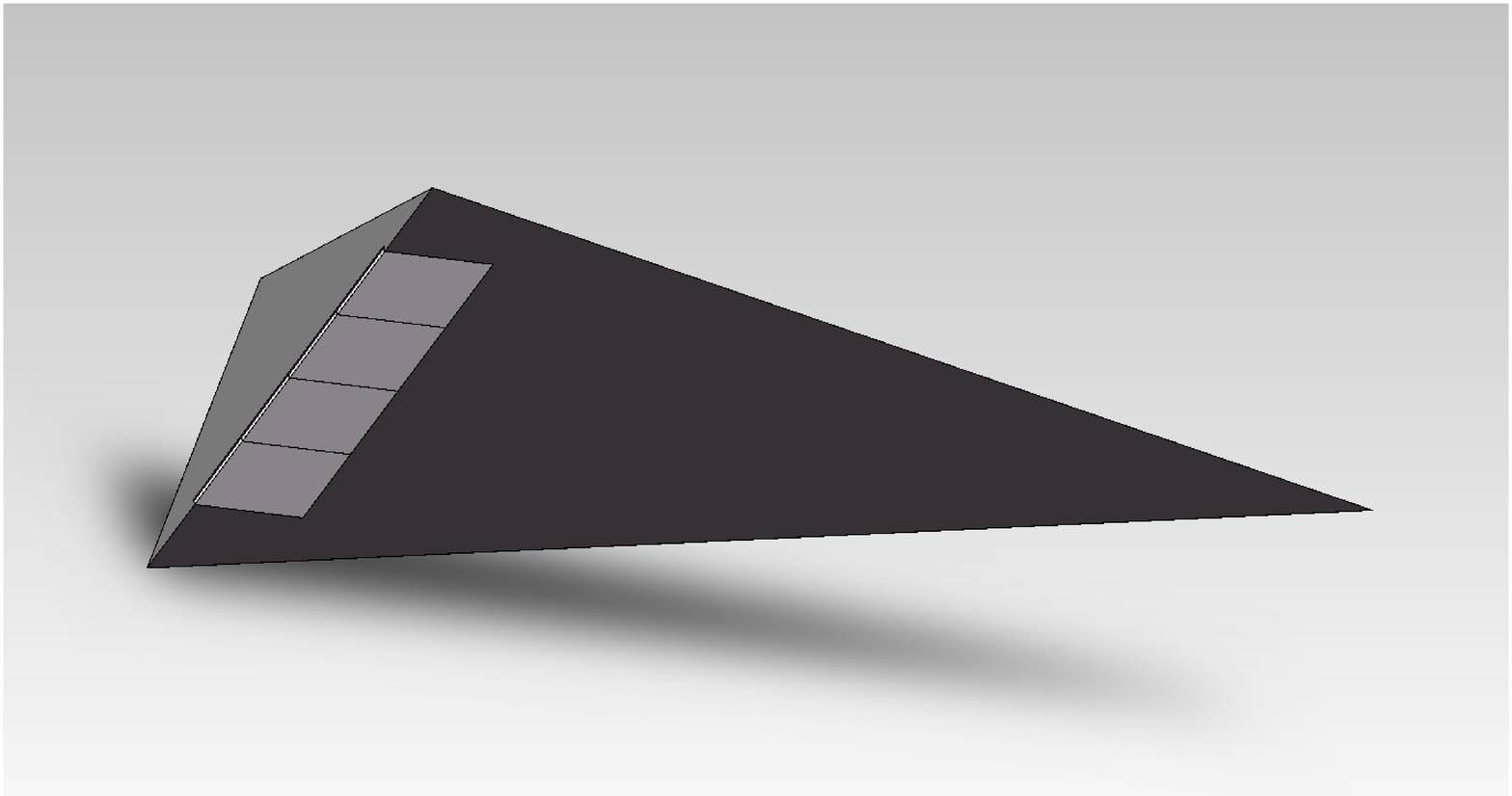}}
     \caption{Hypersonic vehicles, where (a) an online artistic illustration of HTV-2 and (b) the illustration of our imaginary vehicle.}\label{HSVmodel}
     \end{figure}

Very limited experimental results have been documented for surface pressure fluctuations at hypersonic speed. Only some empirical equations have been proposed, based on preliminary experiments and analysis of a straight flat plate at Mach numbers 2, 5 and 8 \cite{coe1972pressure,houbolt1966on,raman1974study}, where compressibility, heat transfer between mediums and viscous effects are incorporated. The equations are quite helpful, summarizing physical insights and setting a good start point for dynamic analysis.  For readers’ convenience, one equation adopted in this work is listed below. It is
\begin{equation}\label{prmse}
p_{rms} = \frac{0.003\rho_e V_e^2}{1+ 0.13M_e^2},
\end{equation}
\noindent where $p_{rms}$ is  the root mean square of pressure fluctuation on a flat plate, $\rho_eV_e^2/2$ is local dynamic pressure and $M_e$ is local Mach number.  The development of the equations was based on experimental data and fluid principles. Details can be found in literatures\cite{laganelli1983prediction}. The above formulation has been justified in various works\cite{laganelli1983prediction}. It is worthwhile to emphasize that the formulation was presented for hypersonic flat plates at small angle of attack. A practical hypersonic vehicle (see Fig. \ref{HSVmodel}) has more complicated shape and generally operates at a large angle of attack. The above formulation is adopted to assist preliminary analysis.

\begin{figure}[!htbp]
     \centering
     \includegraphics[width=140mm]{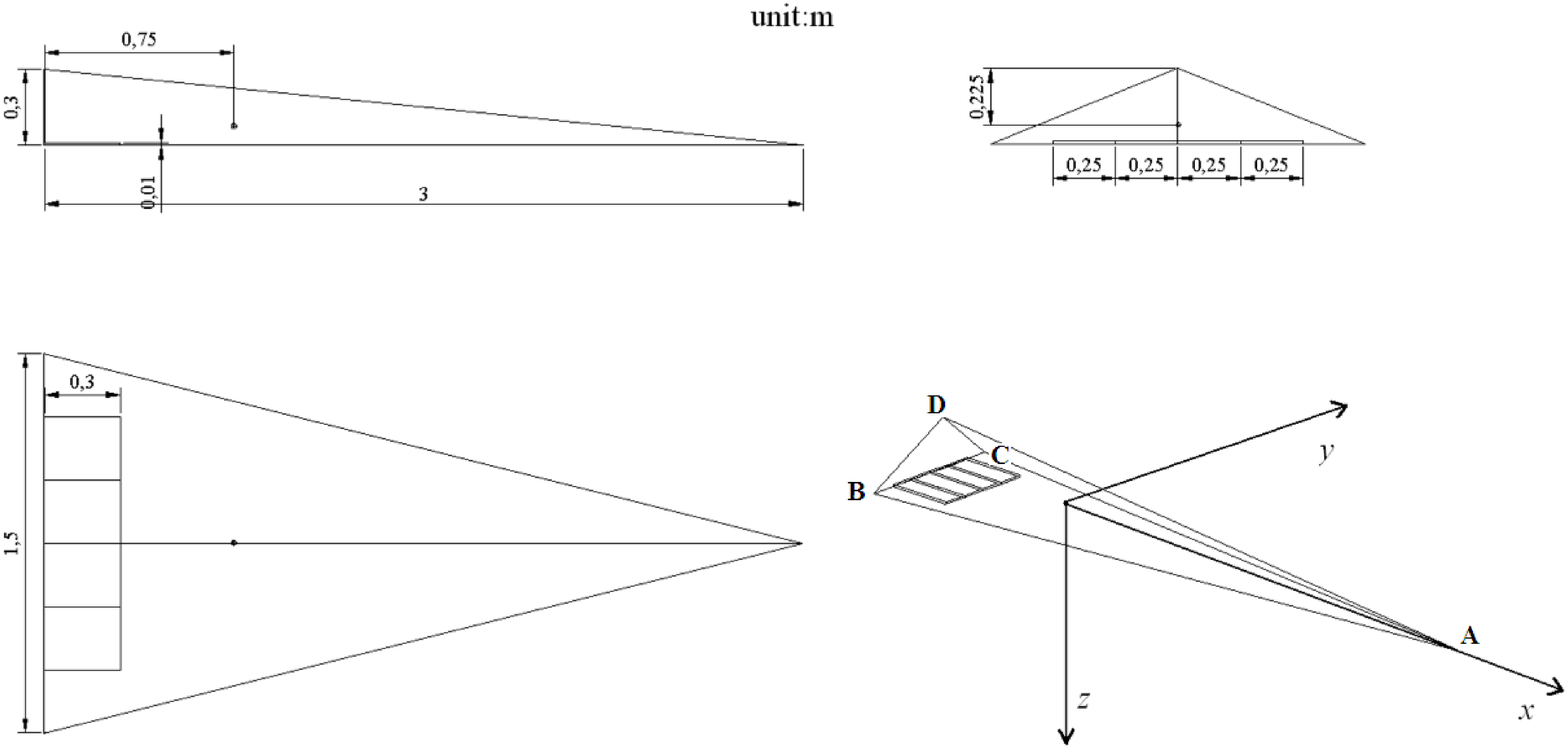}
     \caption{Geometrical specifications of our hypersonic model.}\label{Parameters}
     \end{figure}

It can be seen that surface pressure fluctuations can be estimated using Eq.~(\ref{prmse}) based on the knowledge of local flow field, which can be calculated given the information of aerodynamic profiles.  However, the detailed specifications of HTV-2 are not released to the public. Hence, the following analysis is performed on our own model. Figure \ref{HSVmodel} shows the vehicle, which is drawn according to a beautiful artistic illustration of HTV-2. It can be seen that the cone shapes of HTV-2 are simplified to four flat surfaces (ABC, ABD, ACD, BCD in Fig. \ref{Parameters}) in our model, which ease the following calculations. In addition, four rudders are deployed underneath our model. Two of these rudders control longitudinal dynamics and the other two can control lateral dynamics. This setup decouples roll and pitch actuators and thus simplify the ongoing controller design. The imaginary geometry specifications of our hypersonic vehicle are given in Fig. \ref{Parameters}.  The onboard reaction system is absent on this vehicle. 
     
For simplicity, the mass distribution of the vehicle model is assumed uniform and the density is 2000$\mathrm{kg/m^3}$. 
The moment of inertia is
     \begin{equation}
     I=
     \left[
     \begin{array}{ccc}
     I_{x}  & I_{xy}  & I_{xz}  \\
     I_{yx} & I_{y}  & I_{yz}  \\
     I_{zx} & I_{zy}  & I_{z} 
     \end{array}
     \right]
     =
     \left[
     \begin{array}{ccc}
     24.8675 & 0 & -2.1002 \\
     0 & 548.5759 & 0 \\
     2.1002 & 0 & 395.1200
     \end{array}
     \right].
     \end{equation}
     
Theoretical methods are used to approximate local flowfield of our vehicle. A code in MATLAB is developed to automatically calculate local flowfield based on oblique-shock theory and Prandtl-Meyer theory, given the geometrical shapes, angle of attack ($\alpha$), angle of sideslip ($\beta$) and atmospheric conditions. Tables \ref{table1} and \ref{table2} give approximations of aerodynamic parameters at two operating conditions. The aerodynamic coefficients are nondimensionalized to freestream dynamic pressure. The rolling momentum $L$ and the yaw moment $N$ are also calculated based on the given geometrical specifications.

\begin{table}[!htdp]
\caption{Aerodynamic coefficients at height 20km, $M_\infty$ 10 and $\alpha$ 0deg.}\label{table1}
\begin{center}
\begin{tabular}{|c|c|c|c|c|c|c|c|c|}
\hline
$\beta$(deg) &  $C_x$ &$C_y$ &$C_z$ &$C_l$ &$C_m$ &$C_n$ & L(N$\cdot $m)& N(N$\cdot $m)\\ 
\hline
-5&  -0.0016&  -0.0050&  0.0115&  0.0010&0.0019&0.0009&2574.8780&   2356.2393\\
\hline
-4&  -0.0015&  -0.0040&  0.0108&  0.0008&0.0018&0.0007&2057.2254   &1882.5417\\
\hline
-3&  -0.0015&  -0.0030&  0.0103&  0.0006&0.0017&0.0005&1541.3530 & 1410.4732\\
\hline
-2&  -0.0015&  -0.0020&  0.0100&  0.0004&0.0017&0.0004&1026.8181&    939.6287\\
\hline
-1&  -0.0015&  -0.0010&  0.0097&  0.0002&0.0016&0.0002&513.1847  & 469.6090\\
\hline
 0&  -0.0015&   0.0000&  0.0097&  0.0000&0.0016&0.0000&0.0000     &  0.0000\\
\hline
 1&  -0.0015&  0.0010&  0.0097&  -0.0002&0.0016&-0.0002&-513.1847 &  -469.6090\\
\hline
 2&  -0.0015&  0.0020&  0.0100&  -0.0004&0.0017&-0.0004&-1026.8181 & -939.6287\\
\hline
 3&  -0.0015&  0.0030&  0.0103&  -0.0006&0.0017&-0.0005&-1541.3530  &-1410.4732\\
\hline
 4&  -0.0015&  0.0040&  0.0108&  -0.0008&0.0018&-0.0007&-2057.2254&  -1882.5417\\
\hline
 5&  -0.0016&  0.0050&  0.0115&  -0.0010&0.0019&-0.0009&-2574.8780  &-2356.2393\\
\hline 
\end{tabular}
\end{center}
\end{table}

\begin{table}[!htdp]
\caption{Aerodynamic coefficients at height 50km, $M_\infty$ 10 and $\alpha$ 0deg.}\label{table2}
\begin{center}
\begin{tabular}{|c|c|c|c|c|c|c|c|c|}
\hline
$\beta$(deg) &  $C_x$ &$C_y$ &$C_z$ &$C_l$ &$C_m$ &$C_n$ & L(N$\cdot $m)& N(N$\cdot $m)\\ 
\hline
-5&  -0.0016&  -0.0050&  0.0115&  0.0010&0.0019&0.0009&35.7173  & 32.6845\\
\hline
-4&  -0.0015&  -0.0040&  0.0108&  0.0008&0.0018&0.0007&28.5367  & 26.1136\\
\hline
-3&  -0.0015&  -0.0030&  0.0103&  0.0006&0.0017&0.0005&21.3808  & 19.5653\\
\hline
-2&  -0.0015&  -0.0020&  0.0100&  0.0004&0.0017&0.0004&14.2435  & 13.0340\\
\hline
-1&  -0.0015&  -0.0010&  0.0097&  0.0002&0.0016&0.0002&7.1186    & 6.5142\\
\hline
 0&  -0.0015&   0.0000&  0.0097&  0.0000&0.0016&0.0000&0.0000     &0.0000\\
\hline
 1&  -0.0015&  0.0010&  0.0097&  -0.0002&0.0016&-0.0002&-7.1186   &-6.5142\\
\hline
 2&  -0.0015&  0.0020&  0.0100&  -0.0004&0.0017&-0.0004&-14.2435&-13.0340\\
\hline
 3&  -0.0015&  0.0030&  0.0103&  -0.0006&0.0017&-0.0005&-21.3808&-19.5653\\
\hline
 4&  -0.0015&  0.0040&  0.0108&  -0.0008&0.0018&-0.0007&-28.5367&-26.1136\\
\hline
 5&  -0.0016&  0.0050&  0.0115&  -0.0010&0.0019&-0.0009&-35.7173&-32.6845\\
\hline 
\end{tabular}
\end{center}
\end{table}

Control and stability derivatives can be estimated by perturbing attitudes and rudder setups. For example,  in the case of Table \ref{table1}, roll and yaw moment coefficients with respect to $\beta$ is $L_\beta=513.2\mathrm{N\cdot m/deg}$ and $N_\beta=469.6\mathrm{N\cdot m/deg}$. The coefficients are $L_\beta=7.12\mathrm{N\cdot m/deg}$ and $N_\beta=6.51\mathrm{N\cdot m/deg}$ for the case of Table \ref{table2}. 

     \begin{figure}
     \centering
     \subfigure[]
       { \includegraphics[width=70mm]{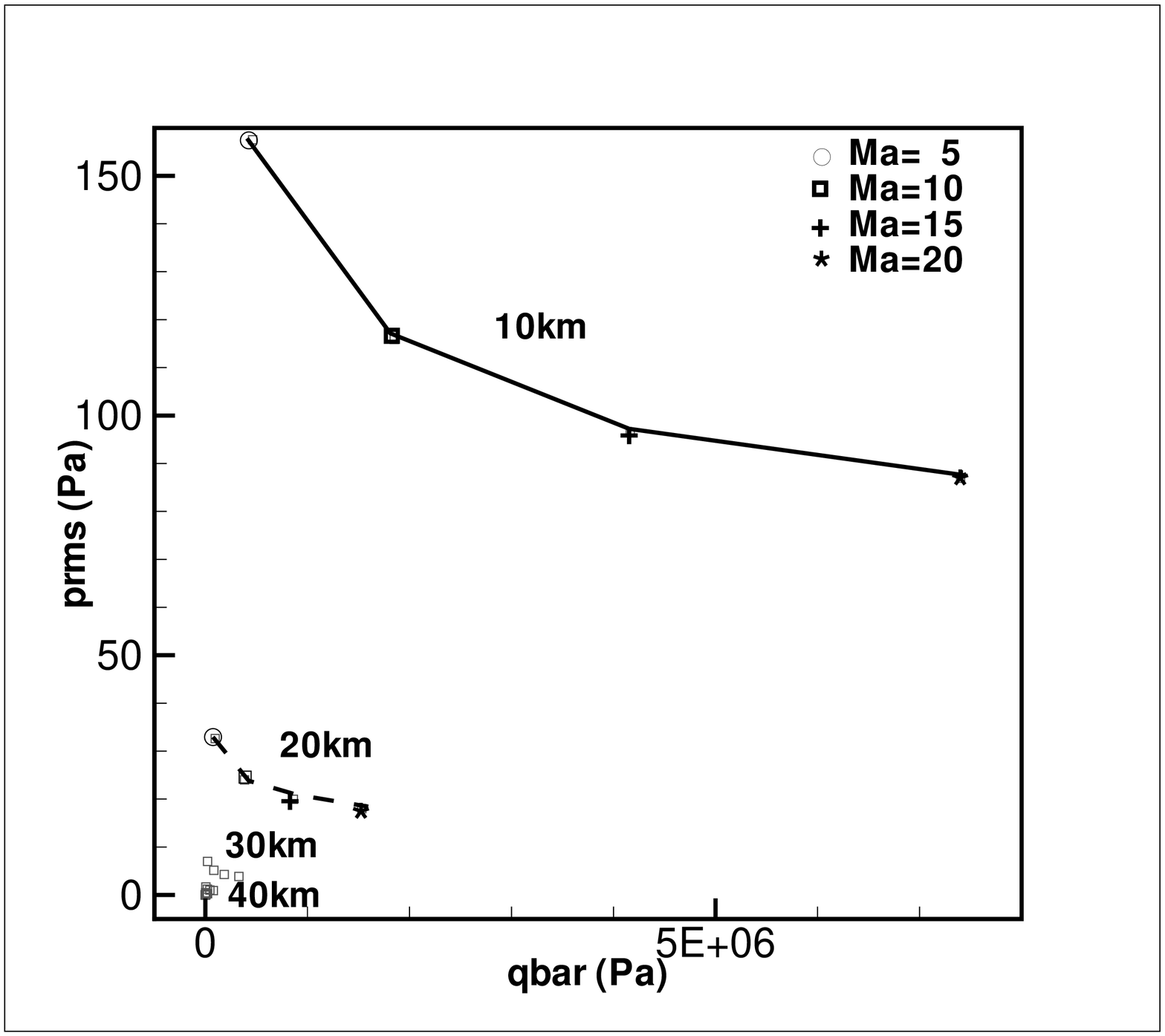}}
     \subfigure[]
       { \includegraphics[width=70mm]{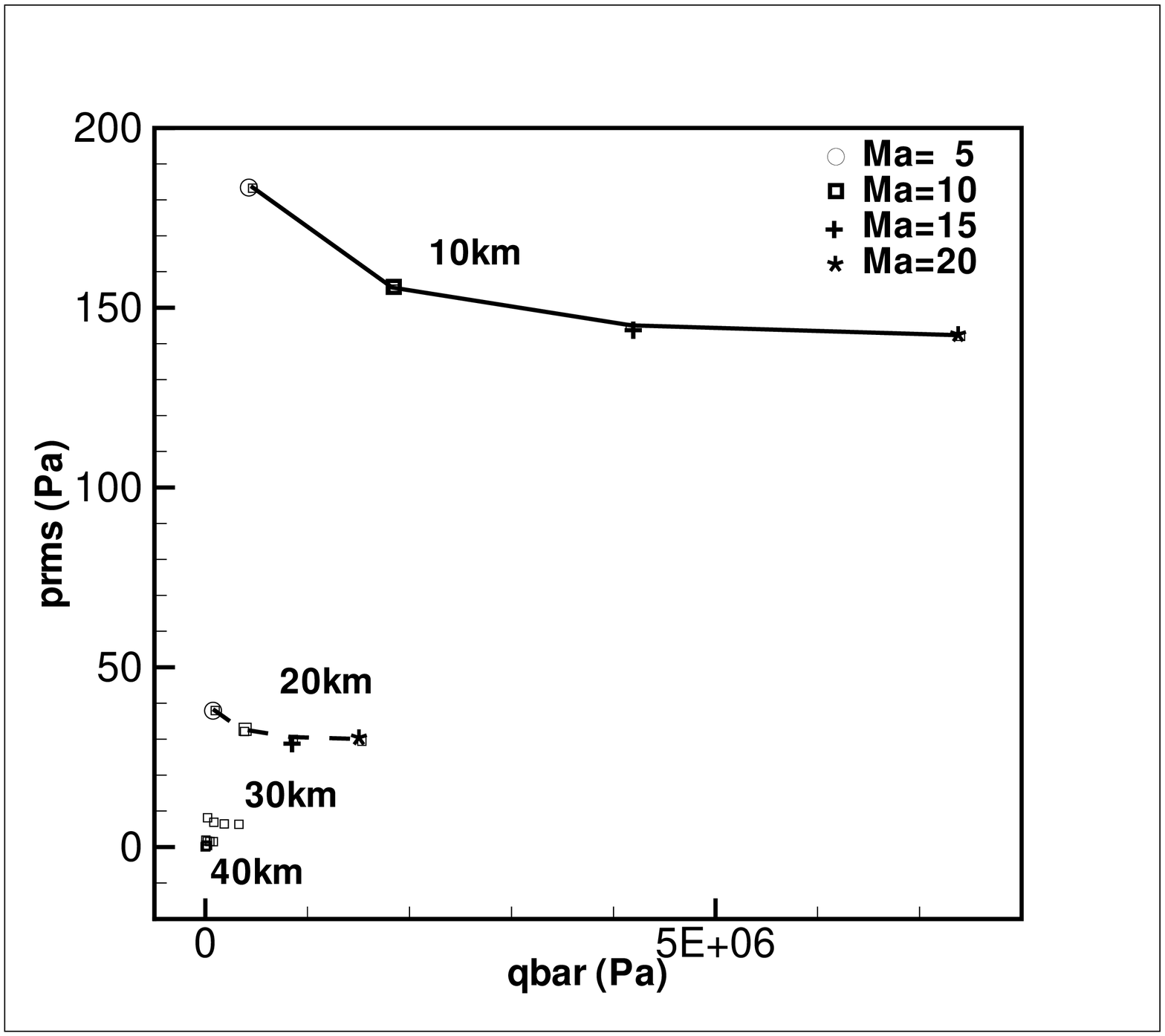}}
     \caption{Relation between $p_{rms}$ and $\bar{q}$ on (a) plane ABD and (b) plane ACD.}\label{prms}
     \end{figure}
Pressure fluctuations on the two lateral surfaces (ABD and ACD) can be calculated using the empirical formula (Eq. (\ref{prms})) based on the fluid approximations. Figure \ref{prms} shows the results, where $\bar{q}$ is dynamic pressure of the freestream. It can be seen that pressure fluctuations rapidly diminish at high altitude. In contrast, the changes of pressure fluctuations are not so sensitive to $\bar{q}$. In addition, $p_{rms}/\bar{q}$ is identical at various heights according to the present formula. In this approximated investigation, the peak $p_{rms}$ is about 140dB, achieved at $M=5$ and 10km height. It was reported in the literature\cite{mixson1987acoustic} that the value of pressure fluctuations on the forward fuselage could reach 160dB, suggesting that the formula adopted here underestimates pressure fluctuations. However, the following analysis shows that even these potentially underestimated fluctuations could cause lateral dynamic instabilities. 

\section{Lateral Stability Analysis}
Five-degree-of-freedom equations of motion are defined relative to body coordinates\cite{cook2007flight}, 
\footnotesize
\begin{eqnarray}
\nonumber
 \dot{p}&=&(\frac{I_y-I_z}{I_x})qr + \frac{I_{xz}}{I_x}\dot{r} + \frac{I_{xz}}{I_x}pq + \frac{Al}{I_x}C_{l\beta}\beta + \frac{Al}{I_x}C_{l\delta_a}\delta_a +  \frac{Al}{I_x}C_{l\delta_r}\delta_r +  \frac{Al^2}{2VI_x}C_{lp}p +   \frac{Al^2}{2VI_x}C_{lr}r,\\
  \dot{q}&=&(\frac{I_z-I_x}{I_y})pr + \frac{I_{xz}}{I_y}r^2 - \frac{I_{xz}}{I_y}p^2 - \frac{I_{x_e}\omega_e}{I_y}r + \frac{A\bar{c}}{I_y}C_{m_\alpha}\Delta\alpha + \frac{A\bar{c}}{I_y}C_{m_{\delta_e}}\delta_e + \frac{A\bar{c}^2}{2VI_y}C_{m_q}q,\\
  \nonumber
 \dot{r}&=&(\frac{I_x-I_y}{I_z})pq + \frac{I_{xz}}{I_z}\dot{p} - \frac{I_{xz}}{I_z}qr + \frac{I_{x_e}\omega_e}{I_z}q + \frac{Al}{I_z}C_{n_\beta}\beta + \frac{Al}{I_z}C_{l\delta_r}\delta_r +  \frac{Al}{I_z}C_{l\delta_a}\delta_a +  \frac{Al^2}{2VI_z}C_{n_r}r +   \frac{Al^2}{2VI_z}C_{n_p}p,\\
 \nonumber
 \dot{\beta} &=& \frac{g}{V}\mathrm{sin}\phi \mathrm{cos}\theta - r\mathrm{cos}\alpha + p\mathrm{sin}\alpha + \frac{A}{mV}C_{Y_\beta}\beta,\\
 \nonumber
  \dot{\alpha} &=& \frac{g}{V}\mathrm{cos}\phi \mathrm{cos}\theta + q - p\mathrm{sin}\alpha - \frac{A}{mV}C_{L_\alpha}\alpha,
\end{eqnarray}
\normalsize
\noindent where $A=\rho V^2S/2$, $S$ is aerodynamic surface area, $\bar{c}$ is the character length, $V$ is speed of freestream, $(p, q, r)$ are  and angular rates, $(\phi, \theta)$ are Euler angles, $l$ is wing span, $\delta_a$ is rudder operating set, $(C_l, C_m, C_n)$ are roll, pitch, yaw momentum coefficients and  $m$ is mass.

The following assumptions can be accepted to simplify the above equations:\\
(1)\ $\alpha$ is time invariant during lateral maneuver, $\alpha=\alpha_0$;\\
(2)\ sin$\alpha \approx \alpha$ and cos$\alpha \approx$ 1; $\phi \approx 0$; \\
(3)\ terms divided by the relatively very large freestream speed ($V$) are neglected;\\
(4)\ $I_{xz}$ can be omitted. 
 
The simplified lateral-directional equation is
\begin{equation}
\dot{p} = L_{\beta}\beta/I_x,\ \dot{r} = N_{\beta}\beta/I_z,\ \dot{\beta }= -r + \alpha_0p.
\end{equation}

It is easy to see
\begin{equation}\label{lateral}
\ddot{\beta}=-\dot{r}+\alpha_0\dot{p}=(-\frac{N_{\beta}}{I_z}+\alpha_0\frac{L_{\beta}}{I_x})\beta.
\end{equation}

The lateral dynamics can be examined by analyzing this equation. For example, inherent oscillation frequency ($\omega_d$) is determined as
\[ \omega_d = i\sqrt{-\frac{N_{\beta}}{I_z}+\alpha_0\frac{L_{\beta}}{I_x}}.\]

It should be noted that the aerodynamic coefficients in Eq. (\ref{lateral}) actually consist of stationary and fluctuation parts, that is
\begin{eqnarray}
\ddot{\beta} &=&(-\frac{N_{\beta_0}}{I_z}+\alpha_0\frac{L_{\beta_0}}{I_x})\beta+(-\frac{N_{\beta_f}}{I_z}+\alpha_0\frac{L_{\beta_f}}{I_x})\mathrm{sin}(\omega_f t)\beta,
\end{eqnarray}
\noindent where the subscript ($0$) denotes stationary parts, and the subscript ($f$) denotes time variant parts whose frequency is $\omega_f$. Without loss of generality, only one single frequency is considered here. The time variant parts can be mainly from the surface pressure fluctuations, which are omitted in classical dynamic analysis for traditional flights. In this work, the above mentioned calculations of our hypersonic model suggest that the stationary aerodynamic coefficients are about one order of magnitude larger than the fluctuating pressure parts (e.g. see Table \ref{table1}, Table \ref{table2} and Figure \ref{prms}).  As a result, we argue that it could be improper to neglect  time variant terms in dynamic equations for hypersonic vehicles. We can take this argument one step further and hypothesize that a controller synthesized based on a dynamic model with the absence of those fluctuating coefficients may lead to HTV-2 test failures. 

To show the idea, the above equation is written in a succinct form
\begin{equation}\label{Mathieu}
\ddot{\beta} = -[a+b\ \mathrm{sin}(\omega_f t)]\beta,
\end{equation}

\noindent which is the so-called Mathieu differential equation \cite{mclachlan1947theory}, where $a=\frac{N_{\beta_0}}{I_z}-\alpha_0\frac{L_{\beta_0}}{I_x}$ and $b=\frac{N_{\beta_f}}{I_z}-\alpha_0\frac{L_{\beta_f}}{I_x}$. A MATLAB code is programmed to find the numerical solutions of Eq. (\ref{Mathieu}). Given initial values of $a$, $b$, $\omega_f$, $\beta|_{t=0}$ and $\dot{\beta}|_{t=0}$, the numerical solutions of $\beta|_{t=T}$ and $\dot{\beta}|_{t=T}$ can be calculated using Runge-Kutta scheme. The stability of the dynamic equation can be examined by checking the value of $k \overset{\underset{\triangle}{}}{=}
 \mathrm{log}_{10}(\beta|_{t=T}/\beta|_{t=0})/T$, where $T$ should be larger than the period of fluctuations (empirically, one order of magnitude). A positive value of $k$ implies instability and vice versa. Contours of $k$ define the stability region. 

     \begin{figure}[!htpb]
     \centering
     \subfigure[Stability region of 1Hz.]
       { \includegraphics[width=70mm]{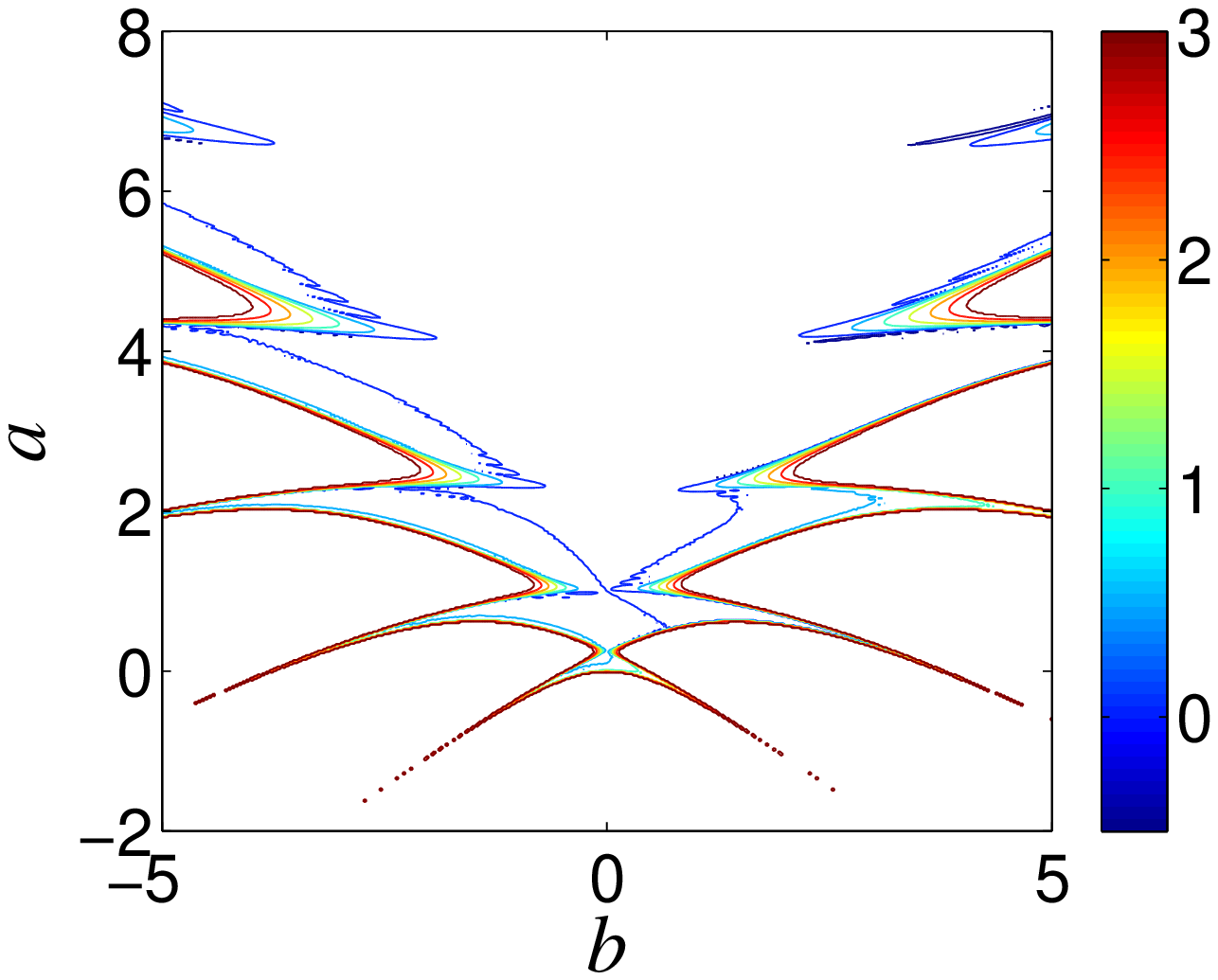}}
     \subfigure[Enlarged figure.]
       { \includegraphics[width=70mm]{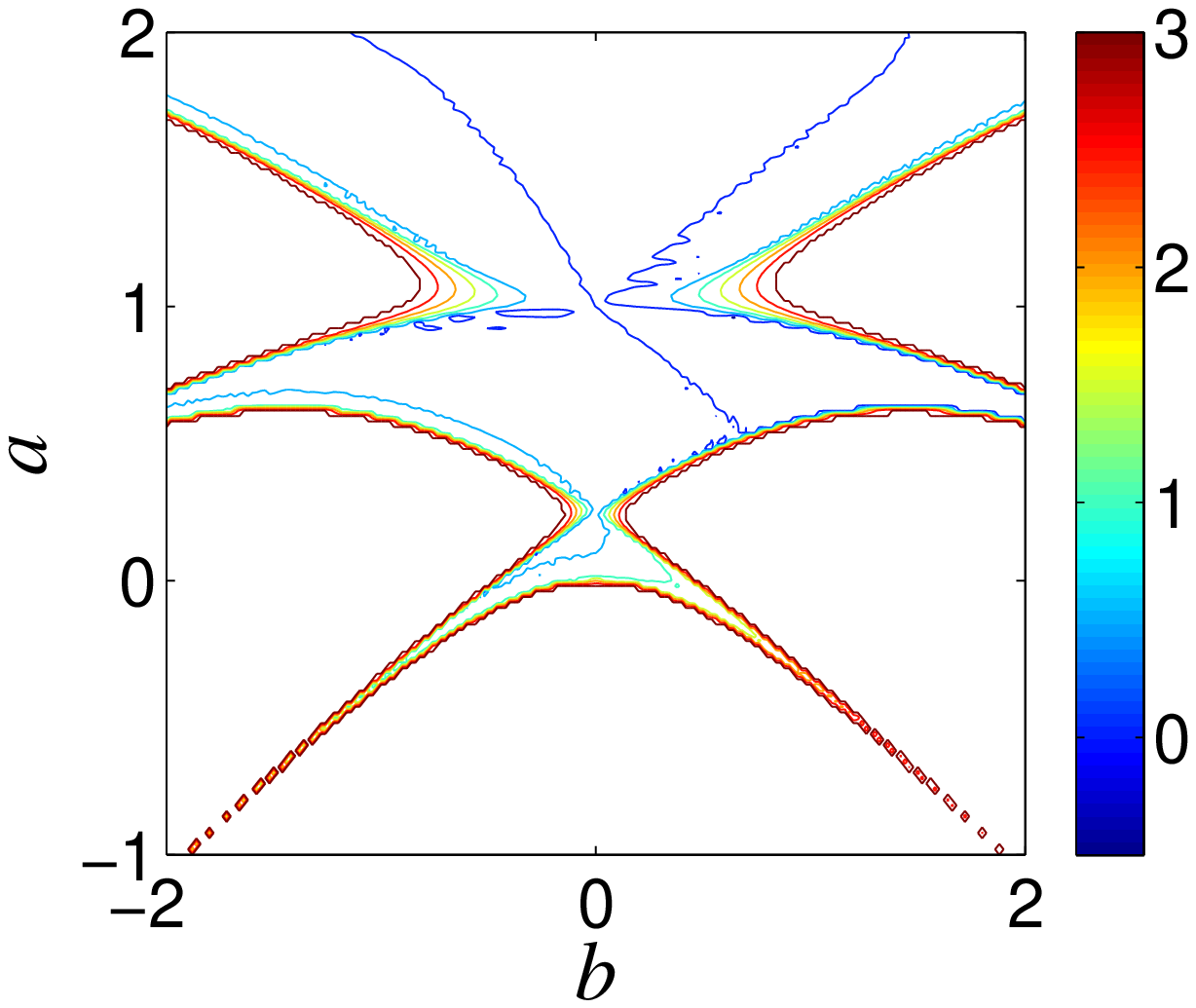}}\\
       \subfigure[Stability region of 10Hz.]
       { \includegraphics[width=70mm]{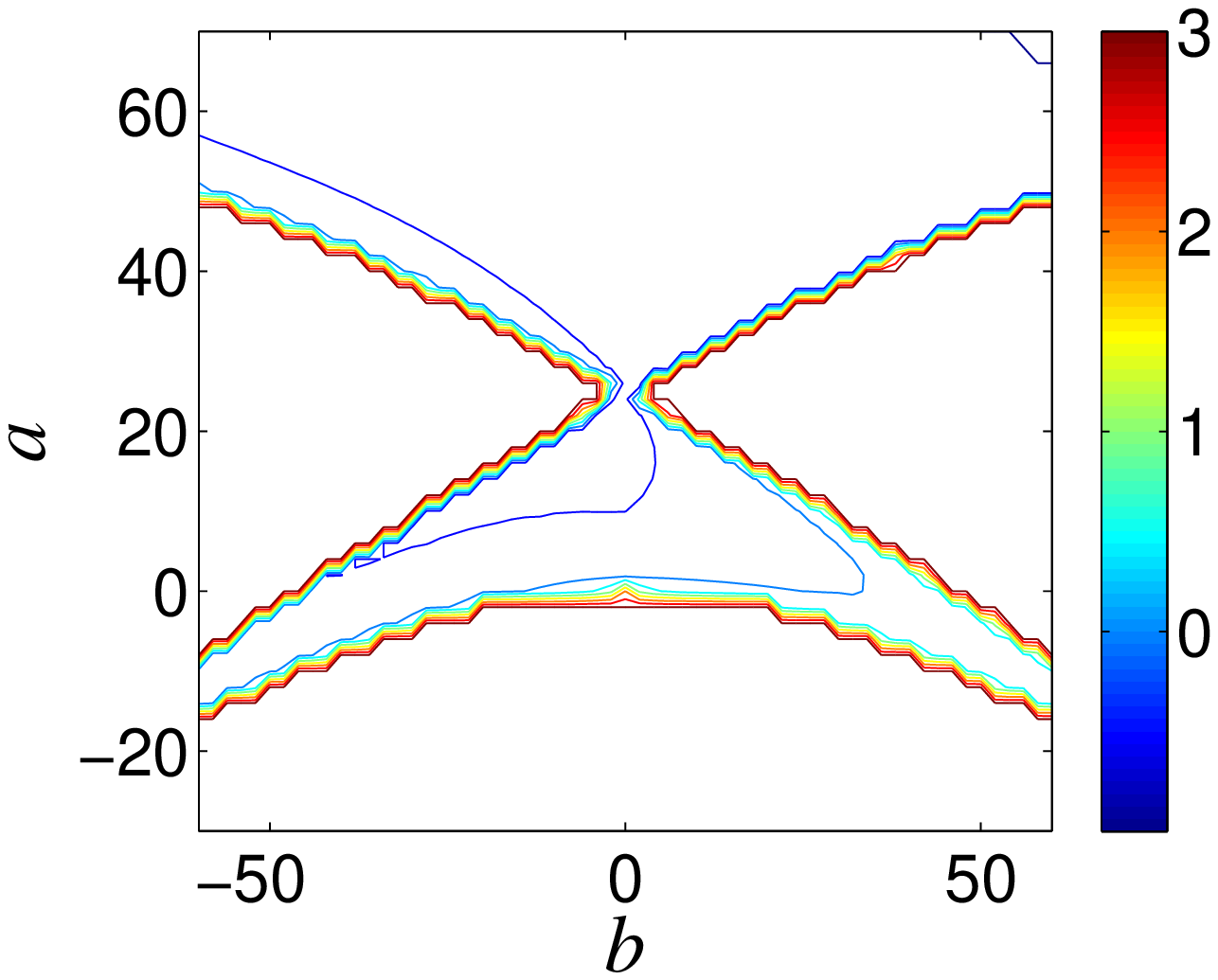}}
     \subfigure[Enlarged figure.]
       { \includegraphics[width=70mm]{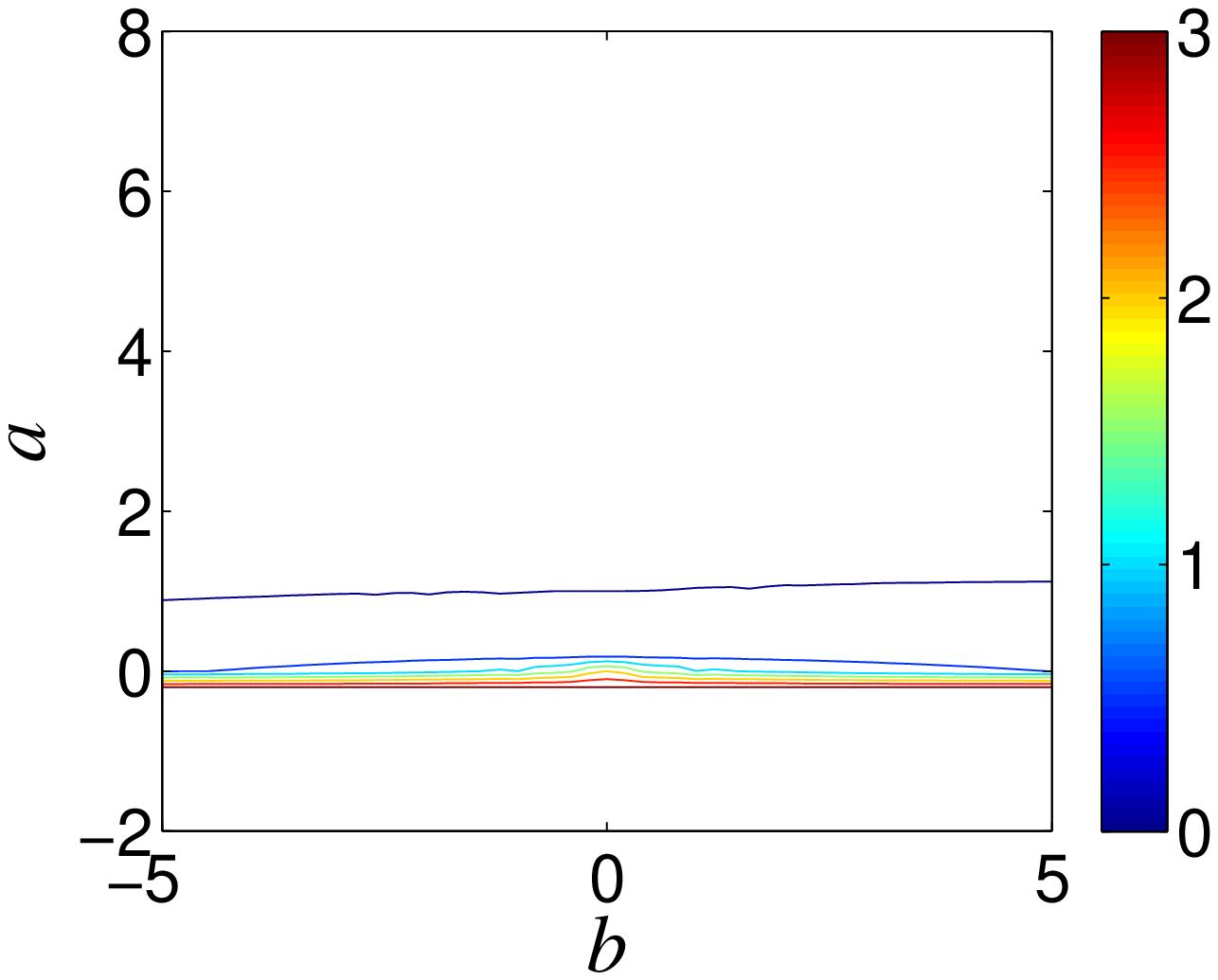}}\\
       \subfigure[Stability region of 100Hz.]
       { \includegraphics[width=70mm]{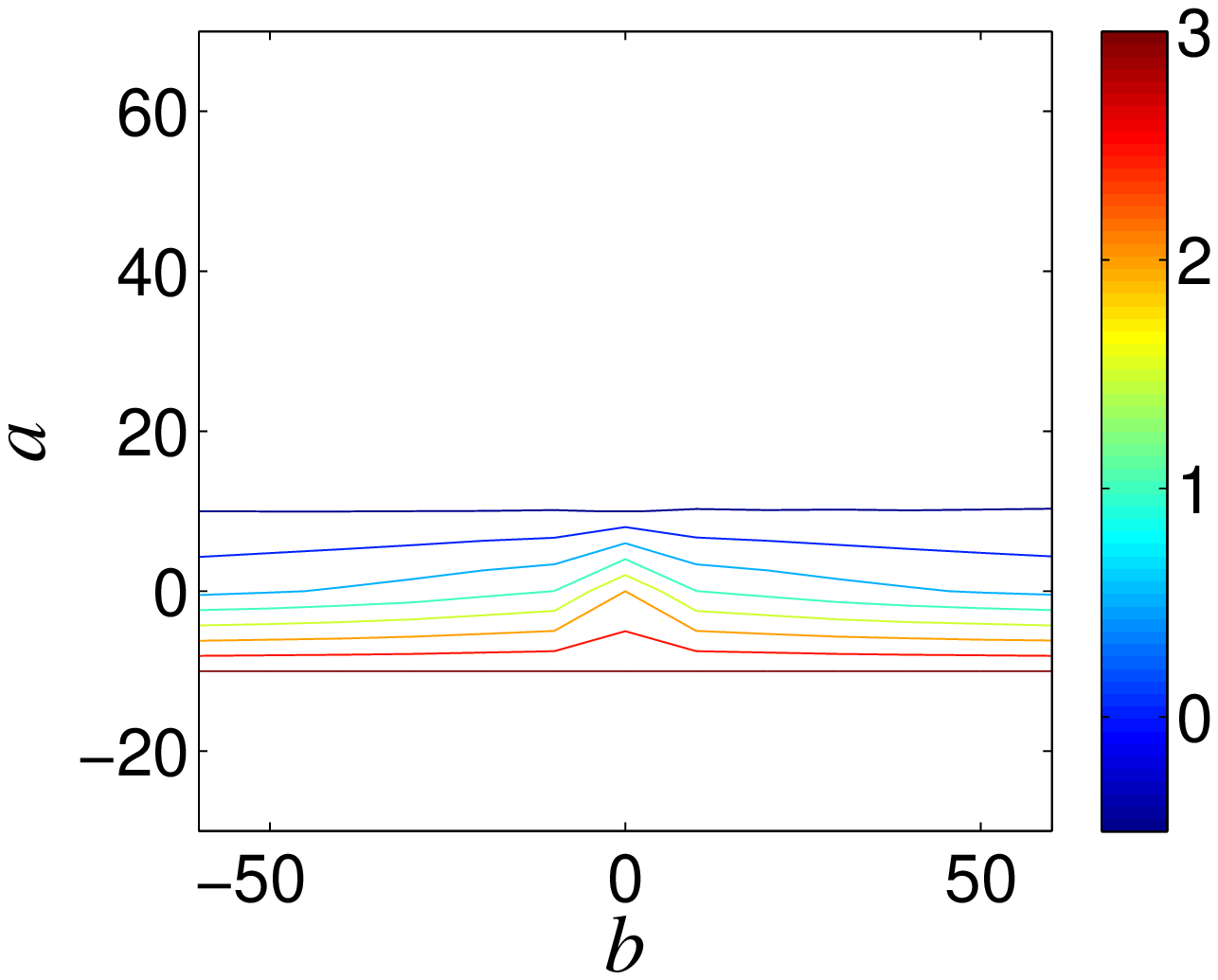}}
     \subfigure[Enlarged figure.]
       { \includegraphics[width=70mm]{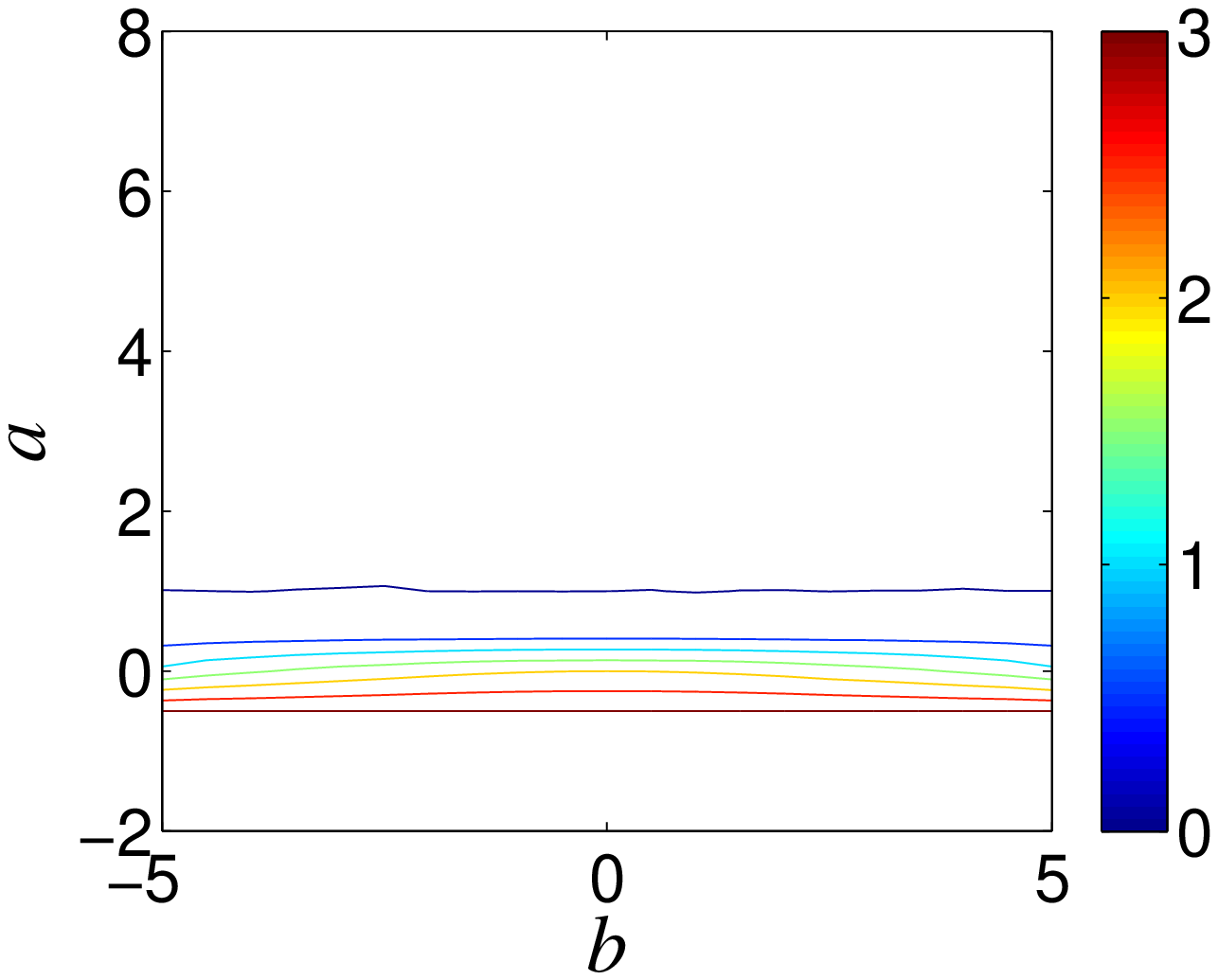}}       
     \caption{Stability region of Eq. (\ref{Mathieu}), where parameter values are given for our flight model. }\label{Mathieustability}
     \end{figure}

Figure \ref{Mathieustability} shows the stability regions at $\omega_f$ =1Hz, 10Hz and 100Hz, respectively. It can be seen that a vehicle can be unstable even if $b=0$, which is the case with the absence of fluctuating terms. It is also obvious, as
negative values of $a$ (and $b=0$) lead the divergence of Eq. (\ref{Mathieu}). The instability can be controlled with the introduction of external actuation. A careful vehicle design can also resolve issue by generating a desired $a$. We can find that exactly the same procedure has been performed after the first test failure of HTV-2. It was said that engineers adjusted center of gravity of the HTV-2 and diminished its angle of attack in the second flight trajectory (to adjust $a$). \cite{DARPA}. In addition, onboard reaction control inputs were enforced (to remedy $a$). Hence, the second test failure of HTV-2 should due to other causes.   Figures \ref{Mathieustability}(a)-(b) show one potential factor. It can be seen that the stability of the system becomes extremely sensitive to the value of $b$, if  the vehicle is suffered 1Hz pressure fluctuations, which can be caused by flow-induced noise and vibrations during hypersonic flight. Once the lateral dynamics becomes unstable, exactly the same phenomenon of HTV-2 flight anomaly  can happen: a higher-than-predicted yawing motion presents, leading to the coupled rolling motion that finally fails the flight . \cite{DARPA}

Figure \ref{Mathieustability} also show some interesting facts. Fluctuations at higher frequencies (10Hz and 100Hz) have little influence on the lateral stability, which counteract our intuition to some extent. It seems a good news as it is well known that the frequency of acoustic perturbation is beyond 20Hz and the frequency of vibration should be larger than 5Hz. However,  Eq. (\ref{Mathieu}) is a nonlinear equation (and so is hypersonic vehicle itself).  High frequency parts can possibly affect low frequency parts as well. The dynamics of  Eq. (\ref{Mathieu}) can become intractably complicated if time variant terms of multi-frequencies are included.  The further analysis is not conducted as we lack the knowledge of the spectrum of pressure fluctuations on hypersonic vehicles.

\section{Summary}
This paper proposes that pressure fluctuations could lead to lateral instability of hypersonic vehicles. A conceptual design imitating HTV-2 to some extent was produced. The related local flowfield was approximated using  oblique-shock wave and Prandtl-Meyer theories. Pressure fluctuations over the vehicle surfaces were derived based on a previous empirical formulation for a hypersonic flat plate. It will be definitely helpful if either a more accurate and expensive computational fluid dynamics can be performed, or an empirical formulation for  a hypersonic wedge is available. In addition, this paper only proposes one potential reason but without a solution. All those constitute ongoing works.  It should be admitted that the initial idea develops from the analysis of HTV-2 anomaly. However, the analytical method developed in this Note should be generic.  

\section{Acknowledgement}
This work has been partially supported by the National Science Foundation Grant of China (90916003) and the Science Foundation of Aeronautics of China (20091571).

\pagebreak
%
%
%
\appendix\label{section:references}
%
%
\bibliographystyle{aiaa}
\bibliography{FPLforHTV2}
%
%
\section{Notation}
\emph{The following symbols are used in this paper:}
\nopagebreak
\par
\begin{tabbing}
  XXX \= \kill
  $A$ \> \hspace{8mm}\hspace{2mm} = \hspace{4mm} $\rho V^2/2$, Pa\\
  $\bar{c}$ \> \hspace{8mm}\hspace{2mm} = \hspace{4mm} the character length, m\\
  $(C_l, C_m, C_n)$ \> \hspace{8mm}\hspace{2mm} = \hspace{4mm} roll, pitch, yaw momentum coefficients \\
  $I$  \> \hspace{8mm}\hspace{2mm} = \hspace{4mm} moment of inertia, $\mathrm{kg\cdot m^2}$\\
  $l$  \> \hspace{8mm}\hspace{2mm} = \hspace{4mm} the wing span, m\\
  $L$ \> \hspace{8mm}\hspace{2mm} = \hspace{4mm} roll moment, $\mathrm{N\cdot m}$\\
  $M$ \> \hspace{8mm}\hspace{2mm} = \hspace{4mm} Mach number\\
  $m$  \> \hspace{8mm}\hspace{2mm} = \hspace{4mm} mass of aircraft, kg\\
  $N$ \> \hspace{8mm}\hspace{2mm} = \hspace{4mm} yaw moment, $\mathrm{N\cdot m}$\\
  $(p,q,r)$  \> \hspace{8mm}\hspace{2mm} = \hspace{4mm} roll, pitch and yaw rate, rad/s\\
  $p_{rms}$ \> \hspace{8mm}\hspace{2mm} = \hspace{4mm} rms fluctuating pressure, Pa\\
  $\bar{q}$ \> \hspace{8mm}\hspace{2mm} = \hspace{4mm} dynamic pressure, Pa\\

  $S$ \> \hspace{8mm}\hspace{2mm} = \hspace{4mm}  aerodynamic surface area, $\mathrm{m^2}$\\
  $(x,y,z)$ \> \hspace{8mm}\hspace{2mm} = \hspace{4mm} coordinates, m\\
  $V$ \> \hspace{8mm}\hspace{2mm} = \hspace{4mm} speed of freestream, m/s\\
  $\alpha$ \> \hspace{8mm}\hspace{2mm} = \hspace{4mm} angle of attack, deg\\
  $\beta$ \> \hspace{8mm}\hspace{2mm} = \hspace{4mm} angle of sideslip, deg\\
  $(\phi, \theta)$  \> \hspace{8mm}\hspace{2mm} = \hspace{4mm} Euler angles, deg\\
  $\rho$  \> \hspace{8mm}\hspace{2mm} = \hspace{4mm} density of air, $\mathrm{kg/m^3}$\\
    
 \textit{Subscripts}\\
  $e$    \> \hspace{8mm}\hspace{2mm} = \hspace{4mm} properties local to aerodynamic shape \\
   $f$  \> \hspace{8mm}\hspace{2mm} = \hspace{4mm} time variant part \\
  $\infty$    \> \hspace{8mm}\hspace{2mm} = \hspace{4mm} freestream conditions\\
  $0$  \> \hspace{8mm}\hspace{2mm} = \hspace{4mm} stationary parts\\

 \end{tabbing}
\end{document}